# Tunable temperature induced magnetization jump in a $GdVO_3$ single crystal

L D Tung*

Department of Physics, University of Warwick,

Coventry CV4 7AL, United Kingdom.

*Abstract*: We report a novel feature of the temperature induced magnetization jump observed along the *a*-axis of the $GdVO_3$ single crystal at temperature $T_M \approx 8$ K. Below $T_M$, the compound shows no coercivity and remanent magnetization indicating a homogenous antiferromagnetic structure. However, we will demonstrate that the magnetic state below $T_M$ is indeed history dependent and it shows up in different jumps in the magnetization only when warming the sample through $T_M$. Such a magnetic memory effect is highly unusual and suggesting different domain arrangements in the supposedly homogenous antiferromagnetic phase of the compound.

* Email address Tung.Le@warwick.ac.uk

## I. INTRODUCTION

The Mott insulating transition metal oxides are among systems with the richest physical properties. In the past decade or so, despite of increasingly worldwide efforts in the studies, understanding many anomalous properties of the Mott insulators is still at infancy. The $RVO_3$ compounds (R = rare earth or Y) are Mott insulators which have some special characteristics. The magnetism of the compounds is driven mainly by the $V^{3+}$ ions which have $d^2$ configuration with two electrons coupled ferromagnetically according to the Hund's rule. The two electrons can occupy two states of the degenerate triplet $t_{2g}$ orbitals, and thus orbital quantum fluctuation (OQF) is expected. When cooling to temperature below $T_{OO}$ (ranging from 141 K to ~200 K for different rare earths [1]), the $RVO_3$ compounds experience the orbital ordering (OO) transition which involves marked redistribution of the valence electron density. OO is usually driven by the Jahn-Teller (JT) lattice distortion which will lift the orbital degeneracy and suppress OQF. Recently, the central discussions on the $RVO_3$ compounds have been focused on OQF versus JT physics [2-5].

The $RVO_3$ compounds have been reported with numerous anomalous magnetic properties including temperature induced magnetization reversal [6-11], low field sensitive character [11,12], staircaselike hysteresis loops [12] etc. In the present paper, we will report a novel feature related to the magnetic memory effect observed along the *a*-axis of the $GdVO_3$ single crystal which have not been detected previously by earlier studies for polycrystalline [13-15] and single crystal sample [1].

## II. EXPERIMENT

$GdVO_3$ single crystal was grown by means of the floating zone technique using a high temperature Xenon arc-furnace. Detail procedure of the crystal growth is similar to compounds with other rare earths which have been previously described [11,12].

Measurements of the *zero*-field-cool (*ZFC*) [Ref.16], field-cooled (FC) magnetization and the magnetic isotherms of the sample were carried out in a commercial SQUID magnetometer. In the FC measurements, the sample is cooled from paramagnetic region to 1.8 K in an applied field. In here, the data can be taken either on cooling (FCC) or on warming (FCW). For the *ZFC* measurements, the sample is cooled in *zero* field to 1.8 K before the magnetic field is applied. The data are then taken on warming. Heat capacity measurements of the sample were carried out in a Physical Property Measurement System (PPMS) using a heat capacity option.

**III. RESULTS**

In Fig. 1, we present the heat capacity C and C/T as a function of temperature for the compound. The OO transition occurs at $T_{OO}$ = 199 K and then is followed by an antiferromagnetic (AF) spin ordering (SO) transition at $T_{SO}$ = 118 K. In the ordering region below $T_{SO}$, we observed another transition at temperature $T_M$ of about 8 K. Our heat capacity data are similar with those previously reported on single crystal sample by Miyasaka et al. [1].

The results of FCC, FCW magnetization of the $GdVO_3$ single crystal along the main axes are presented in Fig. 2. In here the *a*- *b*- and *c*-axes are defined according to the *Pbnm* orthorhombic lattice with the lattice parameters *a* = 5.342 Å, *b* = 5.604 Å, *c* = 7.637 Å [17]. We would like to note that, for $GdVO_3$, the data of the neutron diffractions measurements are not available due to the neutron absorbent nature of Gd. However, for other compounds with R = La-Dy, it is known from powder neutron diffraction results [18] that the magnetic structure is of C-type, i.e. the spins order antiferromagnetically in the *ab*-plane and ferromagnetically along the *c*-axis. Since Gd is in between La and Dy, it is therefore reasonable to assume also the C-type magnetic structure for the compound.

From Fig. 2, it can be seen that the two transitions at $T_{SO}$ and $T_M$ observed from the previous heat capacity data are again shown up in the magnetization data. At the OO temperature, we cannot detect any anomaly in the inversed susceptibility (Fig. 3). The Curie-Weiss behavior is perfectly followed at temperatures few degrees above $T_{SO}$ and the fittings give the effective moment $\mu_{eff}$ = 8.30±0.05 $\mu_B$/f.u. and Weiss temperature $\theta_p$ = -17±1.75 K along all the main axes. Since, in the paramagnetic region, the system consists of the two different non-interacting spins $V^{3+}$ and $Gd^{3+}$, the effective moments can be estimated through the relationship $\mu_{eff} = \sqrt{\mu_{eff}^2(V^{3+}) + \mu_{eff}^2(Gd^{3+})}$ [12]. Assuming that the spins of the $V^{3+}$ and $Gd^{3+}$ are in the ground state with $\mu_{eff}$ ($V^{3+}$) = 2.83 $\mu_B$ (spin only, S = 1) and $\mu_{eff}$ ($Gd^{3+}$) = 7.94 $\mu_B$, an effective moment $\mu_{eff}$ ($GdVO_3$) = 8.43 $\mu_B$ is obtained, which is very close to the observed experimental value. A noteworthy feature in the FCC, FCW data is also on the magnetization reversal observed along the *a*-axis at two temperatures denoted as $T_o$ and $T_s$ in Fig. 2. The values of $T_o$ and $T_s$ are dependent on the applied field and also on whether it is derived from FCC or FCW. At large enough field, e.g. H = 0.5 kOe, there is no magnetization reversal nor any anomaly around $T_s$ and $T_o$ (data not shown).

In Fig. 4, we present the magnetic isotherms of the compounds measured along the main axes at different temperatures. In the "*high*" field regime, we observe a change in the features of the magnetization curves. Below $T_M$, there appear the field induced transition(s) along the *a*- and *b*-axes and the hysteresis along all the main axes which become disappeared at temperatures above $T_M$. The details of the magnetic isotherms around the origin are blown up and displayed in Fig. 5. It can be seen that there is also a characteristic change in the behavior of the magnetic isotherms at temperatures below and above $T_M$. Below $T_M$, we observe no remanent magnetization and coercivity (Fig. 5a). The remanent magnetization and coercivity develop noticeably along the *a*-axis at temperature above $T_M$ (Fig. 5b, c, d) which result in the magnetization reversal observed only along this direction (Fig. 2). The coercivity at 110 K (close to $T_{SO}$) is even higher than that obtained at 50 K and 12 K. The increase of the coercivity with increasing temperature observed in $GdVO_3$ is in contrast with conventional magnets in which thermal energy should lead to the reduction of the coercivity.

Since the *a*-axis is a "*peculiar*" direction, we have carried out in more details with further measurements. Earlier, we have reported that, in many of the $RVO_3$ compounds, the *ZFC* magnetization can be seriously affected by a presence of the inevitable trapped field (TF) in the superconducting magnet of SQUID [11,12]. We have examined the TF carefully. Before each measurement, we ran the degauss sequence to minimize the TF, its absolute value is estimated to be less than 2 Oe. We can "*generate*" the TF with opposite sign just by reversing the sign of the magnetic fields in the degauss sequence. In Fig. 6, we present the *ZFC* data of the compound bound with positive TF (PTF) and negative TF (NTF). From the Figure, it can be seen that the transition temperature $T_M$ separates the *ZFC* curves into two distinctively different behaviors. Below $T_M$, the *ZFC* magnetization always follows the direction of the applied field in which it is being measured (hereafter denoted as $H_{meas}$). The two *ZFC*_PTF and *ZFC*_NTF curves (almost) coincide to each other (inset of Fig. 6). This observation is consistent with the fact that, below $T_{M,}$ the compound does not have coercivity (Fig. 5a). When warming through $T_{M,}$ the magnetization direction is no longer bound to the direction of $H_{meas}$, but to the TF under which the sample had been cooled previously. It means that, in this case, the magnetization direction can memorize its previous state and behaves accordingly to it, i.e. the magnetic memory effect.

In order to explore this feature further, we carry out two different measurement protocols named as PA and PB. In PA, we first cool the sample in either PTF or NTF to 1.8 K. At this temperature, we start to "*train*" the sample by applying different magnetic field up to 50 kOe. The field is then switch to $H_{meas}$ (chosen as 50 or 10 Oe), and the data are taken following the temperature sweep-up. In Fig. 7, we display the results corresponding with cooling in PTF (7a, 7c), and NTF (7b, 7d). In each Figure, the data of the no-training *ZFC* curve bound with an opposite TF are also added for comparison. In the insets of the Figure we have included also expanded views of the data at temperatures below $T_M$ to indicate that, in this region, there is hardly any effect of the TF as well as the training fields on the obtained data. When warming the sample through $T_M$, the training field clearly shows its effect through different jumps in the sign and magnitude of the magnetization. With increasing the training field, we observed systematically that the jump is suppressed to make the two *ZFC*_PTF and *ZFC*_NTF curves to behave towards each other. The jump of the magnetization at $T_M$ in a certain $H_{meas}$, thus, can be controlled through the training field.

In another measurement protocol PB, we first cool the sample in different magnetic fields (i.e. FC). Then at 1.8 K, the magnetic field is switched to $H_{meas}$ (chosen as 100 Oe, 50 Oe or 10 Oe), and the data are taken following the temperature sweep-up. In Fig. 8, we display the results with cooling in positive field (8a, c, e) and negative field (8b, d, f). We note that at a certain $H_{meas}$, the FC_PTF and FC_NTF in Fig. 8 are exactly the same as the *ZFC*_PTF and *ZFC*_NTP with "no-training" in Fig. 7, respectively. We have also included, in the insets of Fig. 8, the extended views of the data below $T_M$, to indicate that, in this temperature region, the different cooling fields hardly have any effect on the obtained results. When warming through $T_M$, it is again obvious that the different cooling fields systematically result in different jumps in the sign and magnitude of the magnetization. In this case, it is clear that the jump of the magnetization at $T_M$ in respect to $H_{meas}$ can be controlled both by the sign and the magnitude of the cooling field.

## IV. DISCUSSIONS

The magnetic memory effect observed in $GdVO_3$ is very puzzling. Since it is only shown up through a magnetization jump at a single transition temperature $T_M$, the feature is distinctively different with other magnetic memory systems including spin glass [19] nanomagnetic particles [20] and phase separated manganites [21]. In the latter cases, aging and rejuvenation effects are usually being involved. Earlier, studies on $GdVO_3$ polycrystalline sample [15] reported a transition at 7.5 K (close to $T_M$ of 8 K in our crystal) which was referred to as the antiferromagnetic Néel temperature of the compound. Recently, Miyasaka et al. [1] reported the heat capacity results for $GdVO_3$ single crystal which also shows three transitions as in our heat capacity data. The transition at $T_M$, however, was not discussed into details in their work.

From the magnetic isotherms at 1.8 K in Fig. 4a, we can see that the magnetization is "*saturated*" above applied field of about 40 kOe with a slight anisotropy in the magnetization values obtained along different axes. The saturation magnetization obtained at 50 kOe is of about 6.7 $\mu_B$/f.u. Since this value is very close to a theoretical value of 7 $\mu_B$ for a $Gd^{3+}$ free ion and much larger than 2 $\mu_B$ for a $V^{3+}$ free ion, it is reasonable to think that all of the Gd moments have been forced parallel at field of 50 kOe and they contribute mostly to the saturation magnetization obtained. The $V^{3+}$ moments, on the other hand, are strongly antiferromagnetically coupled, as revealed by the high value of $T_{SO}$ and the negative Weiss temperature. Thus the applied field, which is at least two orders of magnitude smaller than the V-V AF superexchange interaction, should not have any significant influence on the vanadium moments. This scenario, however, can hardly account for the field induced magnetic phase transitions observed at low temperature for the compound, taking into account the fact that $Gd^{3+}$ has S spin character with no crystalline anisotropy and its behavior under the magnetic field should be relatively simple. Obviously there should be some "extra" factors, e.g. related to $V^{3+}$ magnetism, to account fully for the anomalous behavior of the magnetic isotherms at 1.8 K.

The importance of the $V^{3+}$ magnetism is also being revealed through the magnetic memory effect at $T_M$. The transition at $T_M$, is clearly not just only due to the "*trivial*" AF ordering of the Gd moments. Apart from the magnetic memory effect, we also note that coercivity and remanent magnetization can develop only at temperatures above $T_M$. Previously, we have proposed that these latter features as well as the magnetization reversal should be originated from the inhomogeneous nature of the compound due to OQF [11,12]. According to our model, then the transition at $T_M$ should also be related to the fact that the compound is changed from an inhomogeneous $V^{3+}$ antiferromagnetism to a homogeneous one at lower temperature which apparently does not own any coercivity and remanent magnetization. To let it happen, we would speculate that there should be a (significant) JT lattice distortion caused by the AF ordering of the Gd moments at $T_M$ and so it can suppress the effect of OQF of $V^{3+}$ ions. In here, the question of how come the homogeneous AF phase can memorize its history at $T_M$ remains elusive. In any case, the magnetic memory effect observed in $GdVO_3$ compound would suggest that the magnetic state below $T_M$ is to be characterized by different configurations of the magnetic domains in which each configuration has a specific link to sample's history. Such a different domain configuration in homogenous AF phase depending on the history of the sample was observed previously for $YMnO_3$ [22] even though with no such of the magnetic memory effect as in the case of $GdVO_3$.

## V. CONCLUSIONS

In summary, we have studied the magnetic properties of the $GdVO_3$ single crystal. The compound has been shown to exhibit very rich magnetic properties including low field sensitive character, field induced phase transitions, magnetization reversal and magnetic memory effect. The latter is a unique feature of $GdVO_3$ among other orthovanadate compounds and it is distinctively different with many other magnetic memory systems such as spin glass, magnetic nano particles and the phase separated maganites. The controllable magnetization switch in both direction and magnitude may make the compound a potential material for constructing some spin valve devices. We have suggested that the compound is homogenously AF at low temperature but it can have different configurations of domains which show up differently through the magnetization jump at $T_M$. Further experiments, e.g. neutron diffraction to determine the magnetic structure, optical spectroscopy to visualise the domains, as well as theoretical consideration are needed to shed light on this interesting magnetic memory phenomenon.

## ACKNOWLEDGEMENTS

We gratefully acknowledge support from EPSRC, UK.

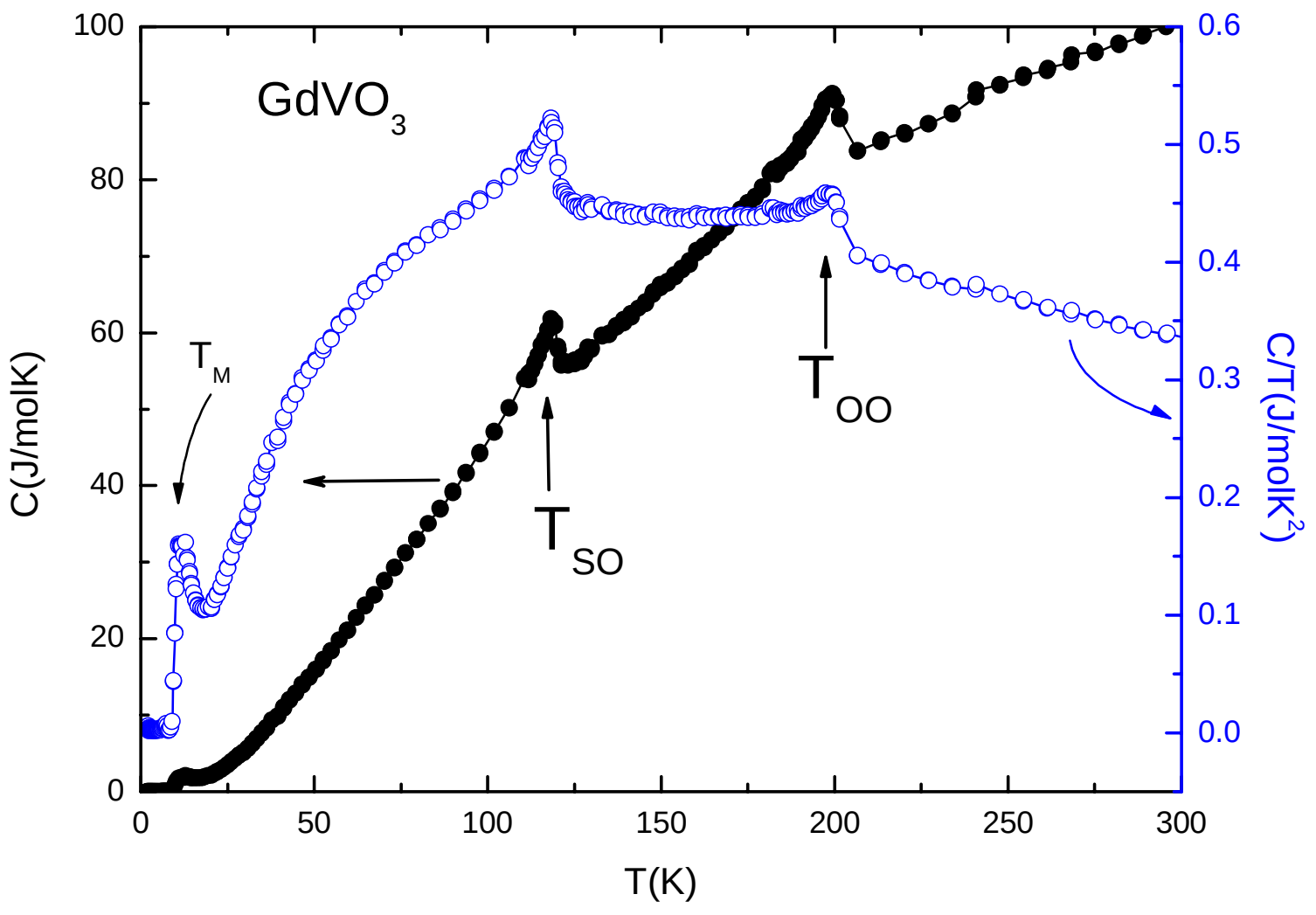

Fig. 1 (colors online): Heat capacity C and C/T as a function of temperature of the $GdVO_3$ single crystal.

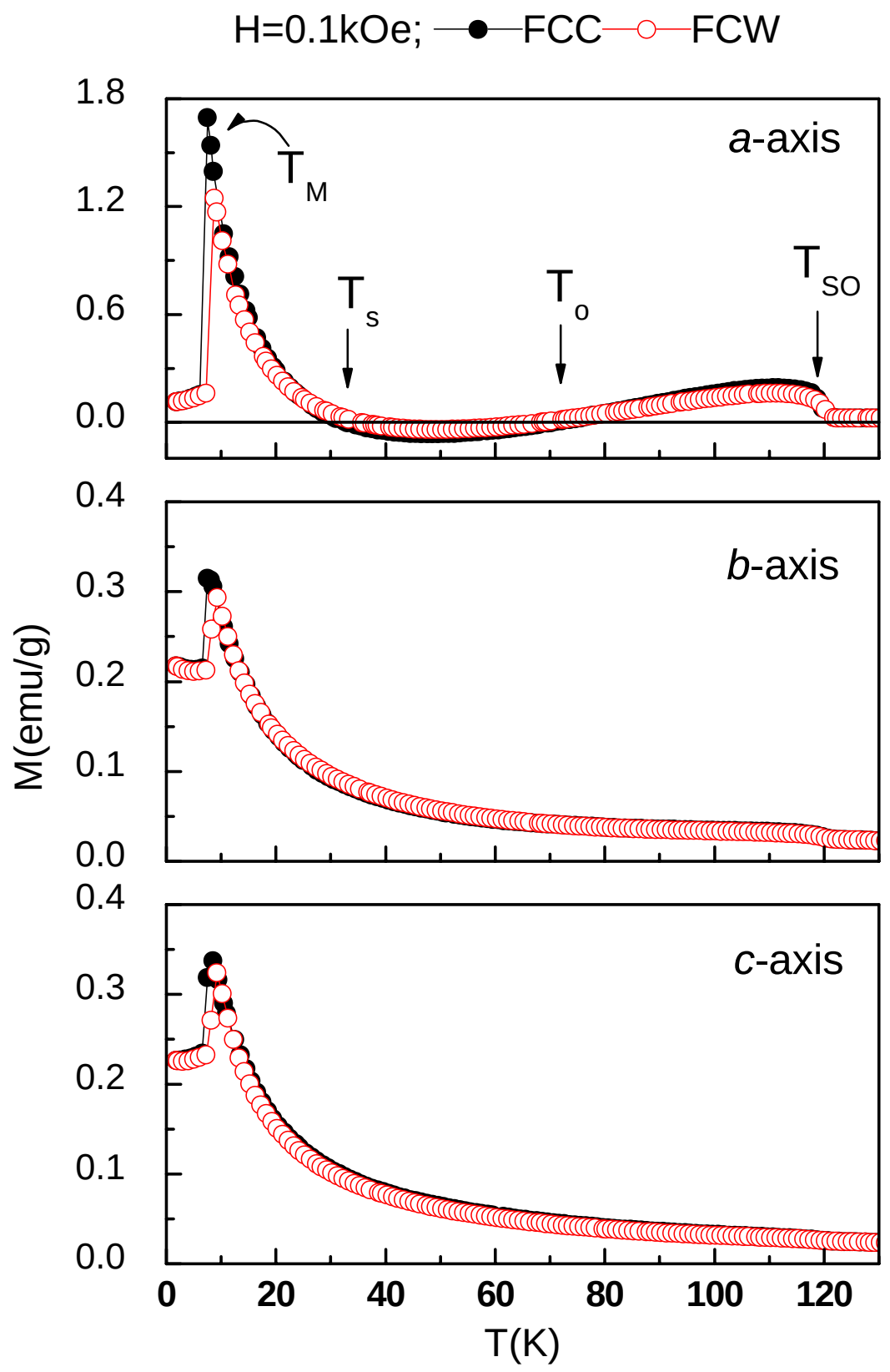

Fig. 2 (colors online): FCC, FCW magnetization of the $GdVO_3$ single crystal measured at 0.1 kOe along the main axes.

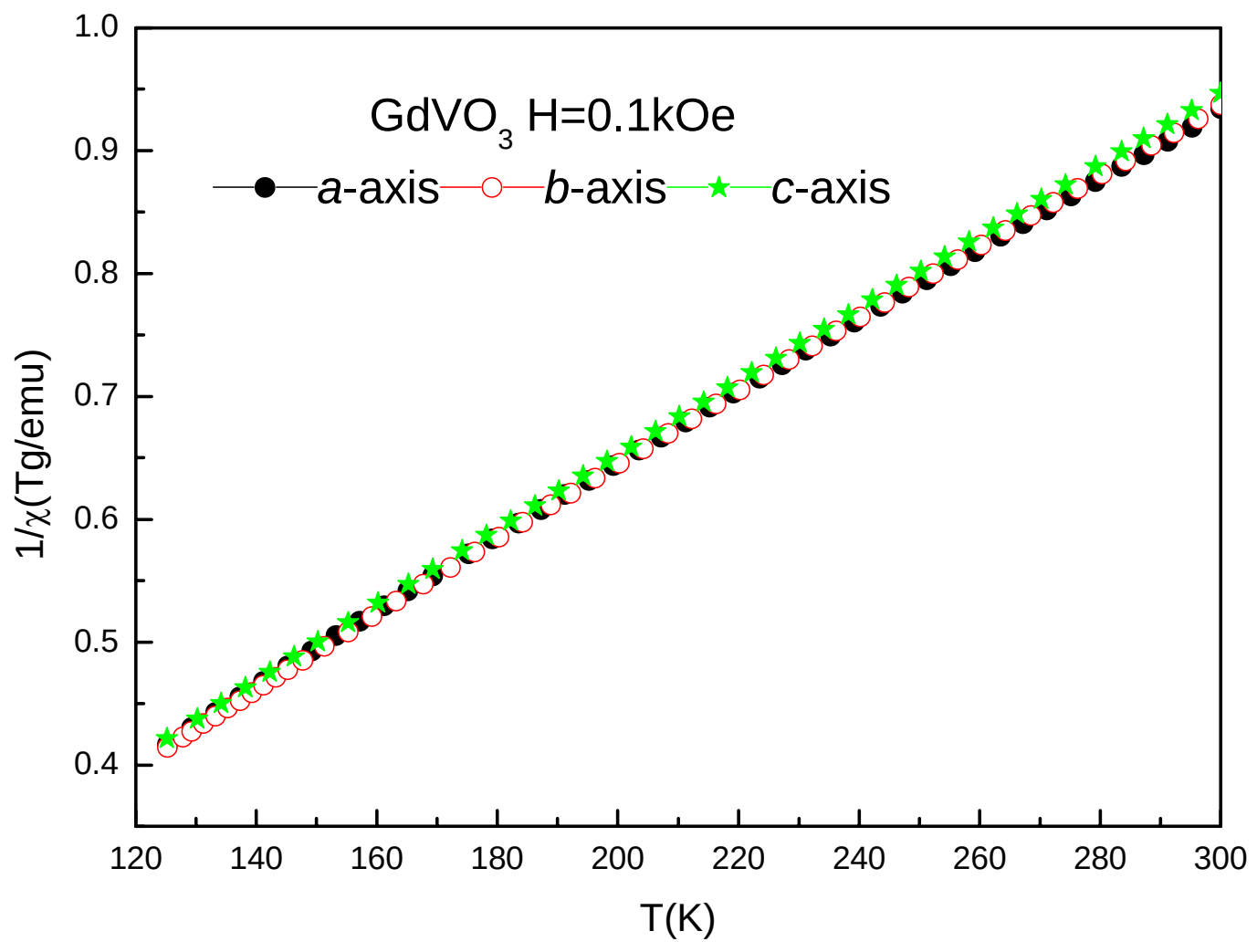

Fig. 3 (colors online): Temperature dependence of the inverse susceptibility in the paramagnetic region of the $GdVO_3$ single crystal.

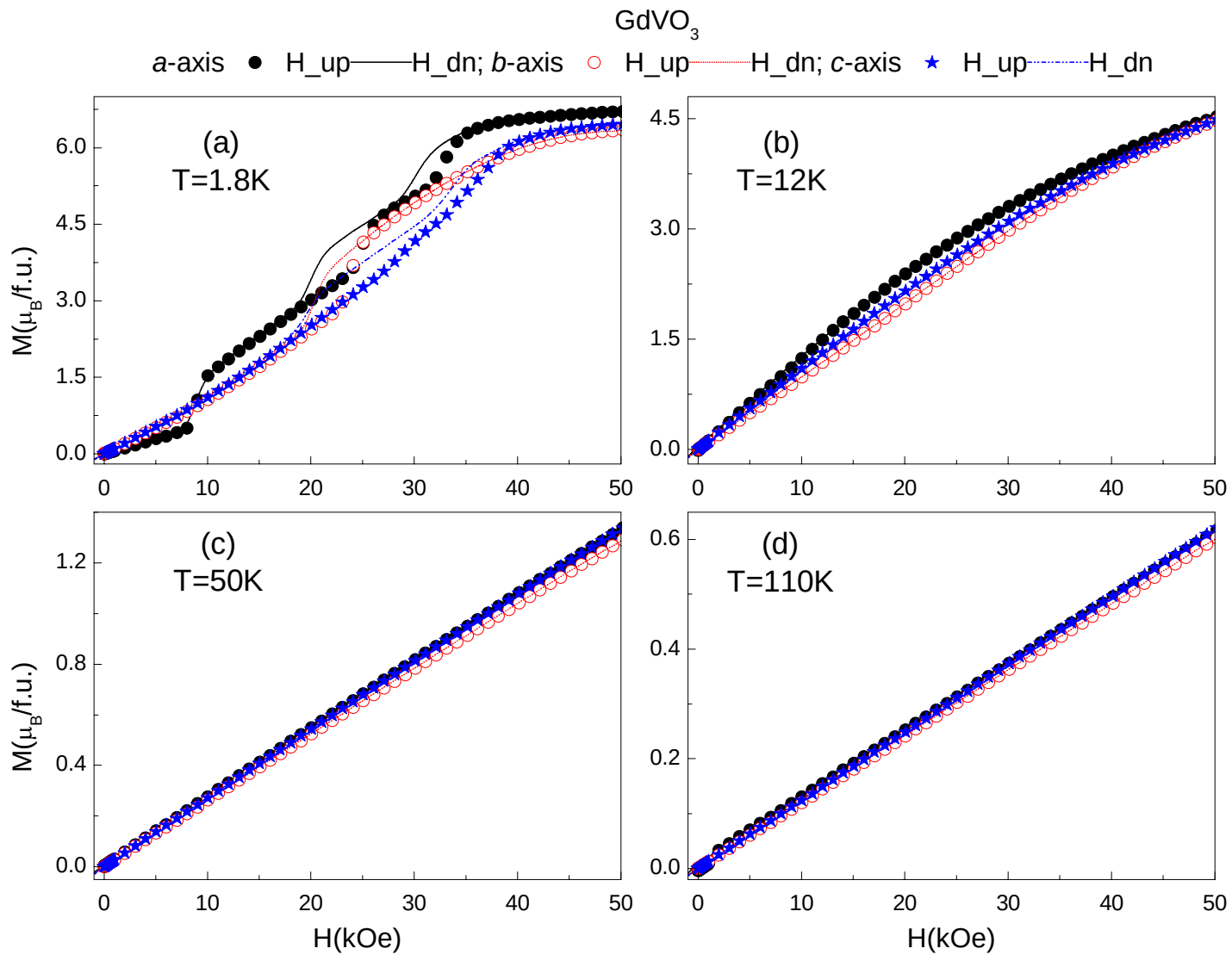

Fig. 4 (colors online): The magnetic isotherms of the $GdVO_3$ single crystal measured along the main axes at different temperatures: 1.8 K (a); 12 K (b); 50 K (c) and 110 K (d). The symbols correspond with increasing field sections, lines with decreasing sections.

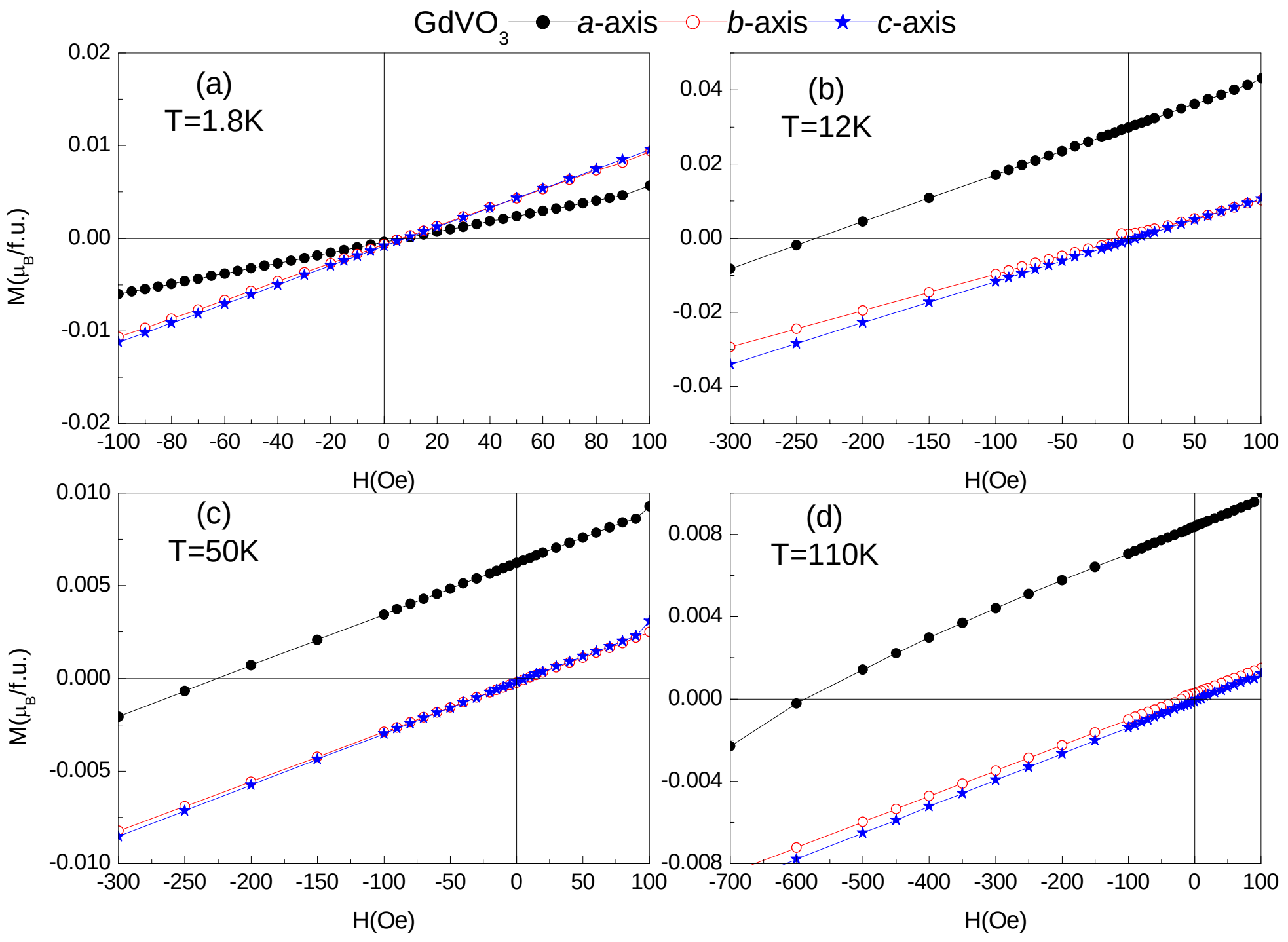

Fig. 5 (colors online): The expanded views around the origin of the decreasing field sections of the magnetic isotherms of the $GdVO_3$ single crystal measured along the main axes at different temperatures: 1.8 K (a); 12 K (b); 50 K (c) and 110 K (d).

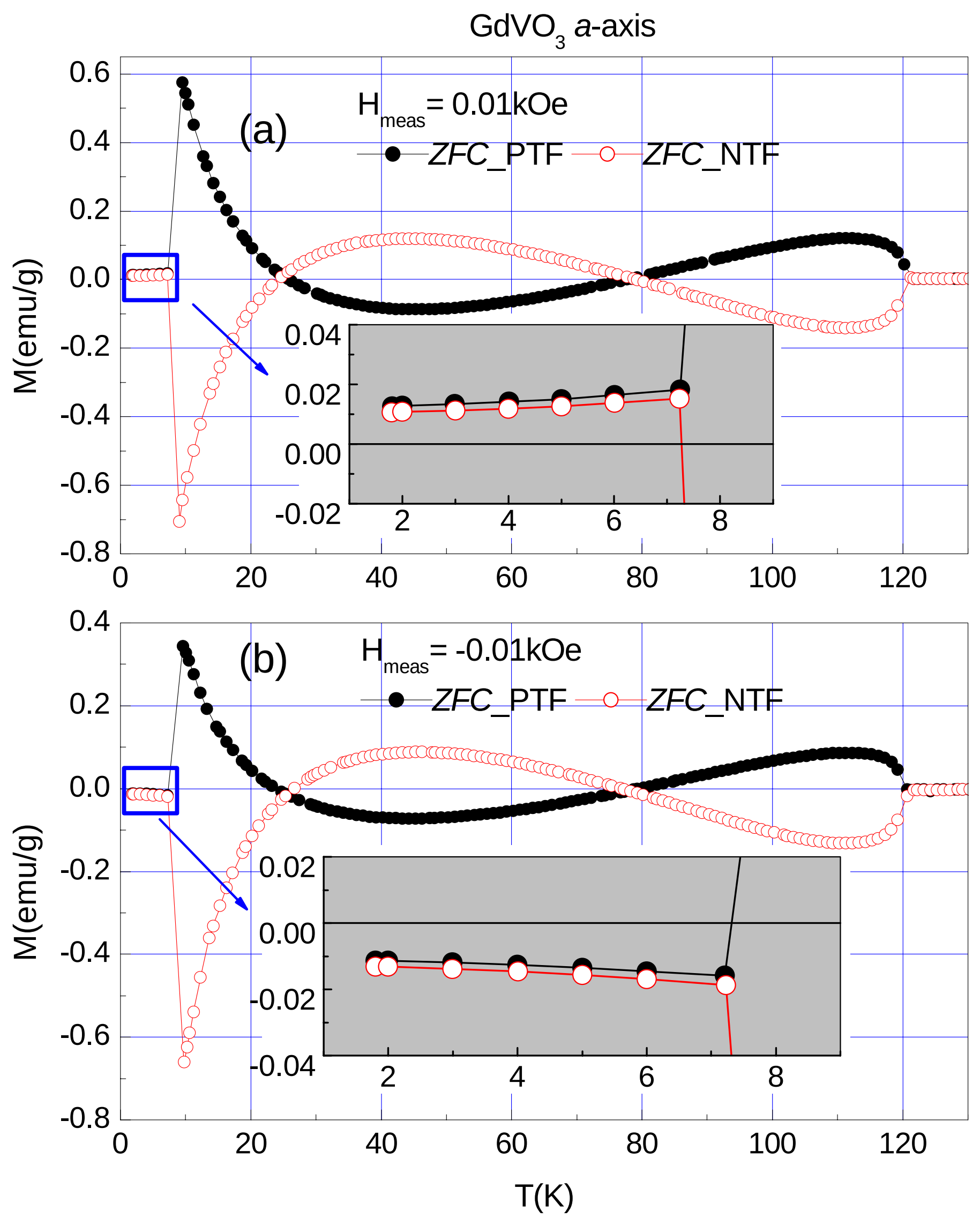

Fig. 6 (colors online): Effect of the trapped field on the *ZFC* magnetization measured along the *a*-axis of the $GdVO_3$ single crystal in applied field $H_{meas}$ of 10 Oe (a) and -10 Oe (b). The inset shows an expanded view in the temperature region below $T_M$.

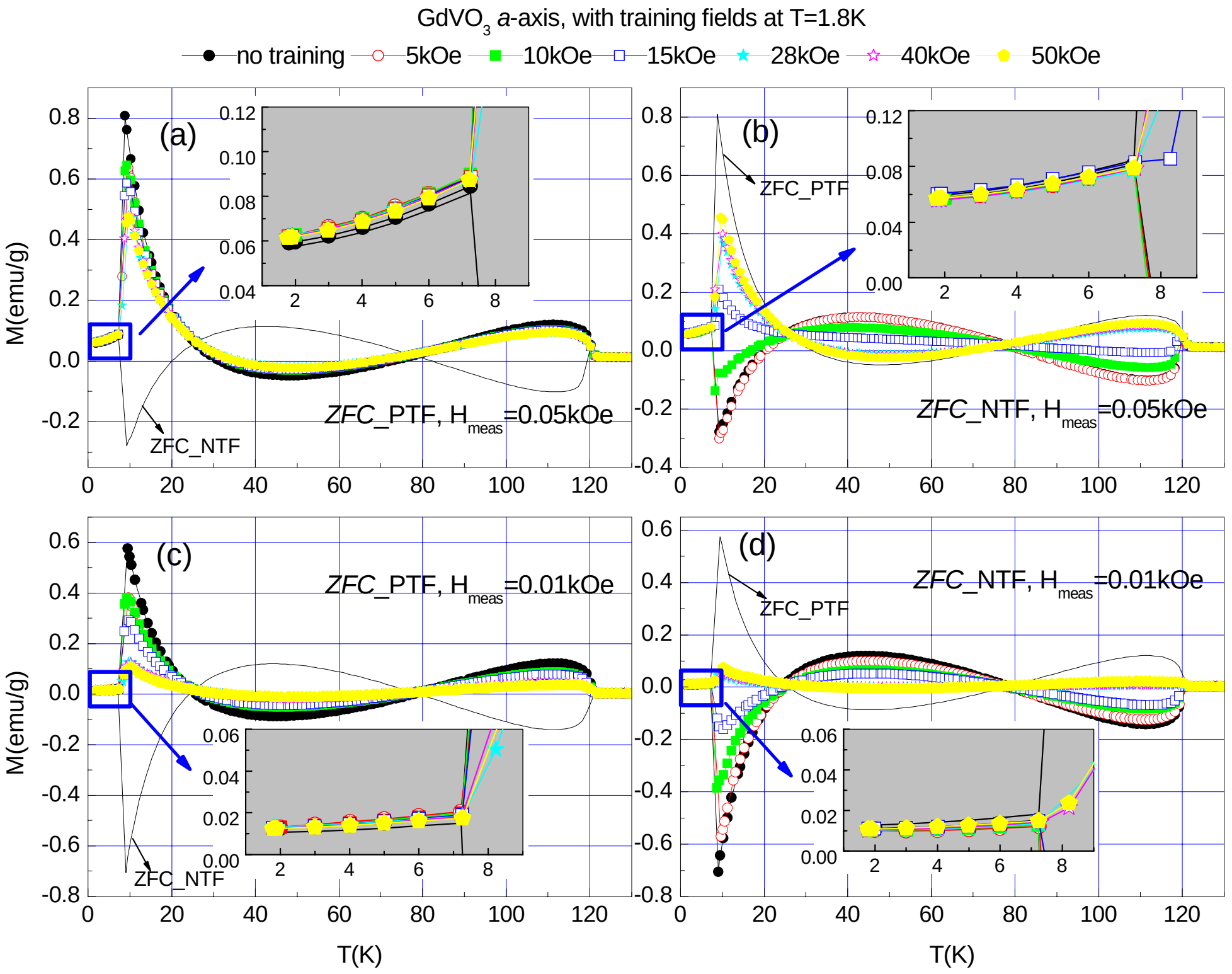

Fig. 7 (colors online): Effect of the training fields at 1.8 K on the temperature dependence of the *ZFC*_PTF (left panels) and *ZFC*_NTF (right panels) measured in different applied fields $H_{meas}$ of 50 Oe (top), 10 Oe (bottom) along the *a*-axis of the $GdVO_3$ single crystal. The inset shows an expanded view in the temperature region below $T_M$.

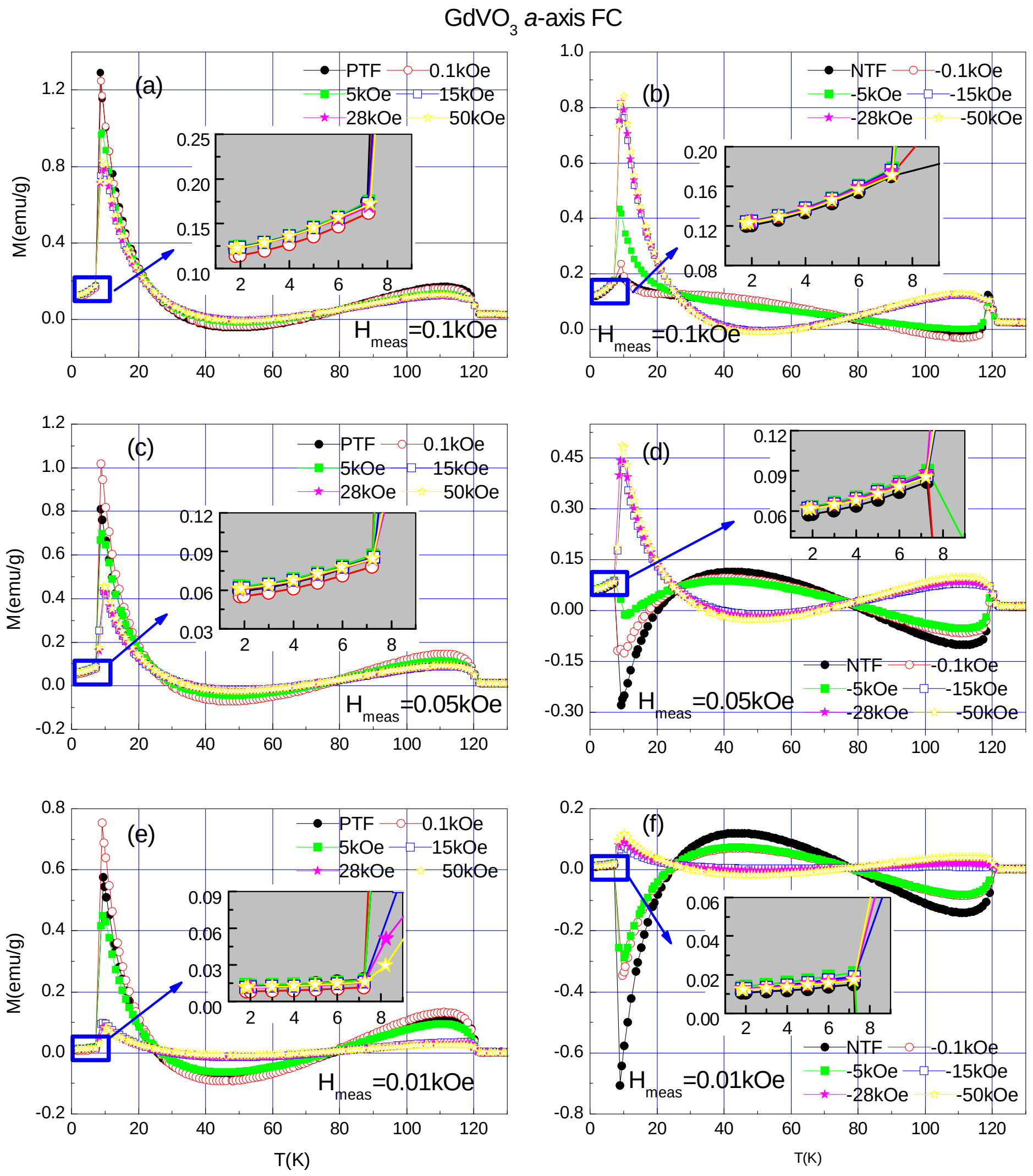

Fig. 8 (colors online): Effect of different cooling fields (positive fields on the left panel, negative fields on the right panel) on the temperature dependence of the magnetization measured in different applied fields $H_{meas}$ of 100 Oe (top), 50 Oe (middle), 10 Oe (bottom) along the *a*-axis of the $GdVO_3$ single crystal. The inset shows an expanded view in the temperature region below $T_M$.